\documentclass[
reprint,
twocolumn,
english,
aps,
prb,
superscriptaddress,
bibnotes,
amsmath,
amssymb,
floatfix,
longbibliography 
]{revtex4-1}

\usepackage[utf8]{inputenc}
\usepackage[T1]{fontenc}
\usepackage{graphicx}
\usepackage{dcolumn}
\usepackage{siunitx}
\usepackage[
    defaultcolor=red, 
    final
]{changes}

\usepackage[colorlinks=true, allcolors=blue]{hyperref}

\graphicspath{{figures/}}

\begin{document}

\title{A chip-scale second-harmonic source via self-injection-locked all-optical poling}

\author{Marco Clementi}
\email{marco.clementi@epfl.ch}
\affiliation{Photonic Systems Laboratory (PHOSL), \'{E}cole Polytechnique F\'{e}d\'{e}rale de Lausanne,  1015 Lausanne, Switzerland}
\author{Edgars Nitiss}
\affiliation{Photonic Systems Laboratory (PHOSL), \'{E}cole Polytechnique F\'{e}d\'{e}rale de Lausanne,  1015 Lausanne, Switzerland}
\author{Junqiu Liu}
\affiliation{Laboratory of Photonics and Quantum Measurements (LPQM), \'{E}cole Polytechnique F\'{e}d\'{e}rale de Lausanne,  1015 Lausanne, Switzerland}
\author{Elena Durán-Valdeiglesias}
\affiliation{Almae Technologies, Route de Nozay, 91460 Marcoussis, France}
\author{Sofiane Belahsene}
\affiliation{Almae Technologies, Route de Nozay, 91460 Marcoussis, France}
\author{Hélène Debrégeas}
\affiliation{Almae Technologies, Route de Nozay, 91460 Marcoussis, France}
\author{Tobias J. Kippenberg}
\affiliation{Laboratory of Photonics and Quantum Measurements (LPQM), \'{E}cole Polytechnique F\'{e}d\'{e}rale de Lausanne,  1015 Lausanne, Switzerland}
\author{Camille-Sophie Brès}
\email{camille.bres@epfl.ch}
\affiliation{Photonic Systems Laboratory (PHOSL), \'{E}cole Polytechnique F\'{e}d\'{e}rale de Lausanne,  1015 Lausanne, Switzerland}

\date{\today}

\begin{abstract}
\noindent
Second-harmonic generation allows for coherently bridging distant regions of the optical spectrum, with applications ranging from laser technology to self-referencing of frequency combs.
However, accessing the nonlinear response of a medium typically requires high-power bulk sources, specific nonlinear crystals, and complex optical setups, hindering the path toward large-scale integration.
Here we address all of these issues by engineering a chip-scale second-harmonic (SH) source based on the frequency doubling of a semiconductor laser self-injection-locked to a silicon nitride microresonator.
The injection-locking mechanism, combined with a high-Q microresonator, results in an ultra-narrow intrinsic linewidth at the fundamental harmonic frequency as small as 41~Hz.
Owing to the extreme resonant field enhancement, quasi-phase-matched second-order nonlinearity is photoinduced through the coherent photogalvanic effect and the high coherence is mapped on the generated SH field. 
We show how such optical poling technique can be engineered to provide efficient SH generation across the whole C and L telecom bands, in a reconfigurable fashion, overcoming the need for poling electrodes.
Our device operates with milliwatt-level pumping and outputs SH power exceeding 2~mW, for an efficiency as high as 280~\%/W under electrical driving.
Our findings suggest that standalone, highly-coherent, and efficient SH sources can be integrated in current silicon nitride photonics, unlocking the potential of $\chi^{(2)}$ processes in the next generation of integrated photonic devices.
\end{abstract}

\maketitle

\section*{Introduction}
\noindent Second-harmonic generation (SHG)\cite{Franken1961} plays a fundamental role in the realm of nonlinear optics, as it enables linking octave-spaced regions of the spectrum while preserving the coherence of the optical field.
Applications range from laser physics and technology\cite{svelto2010principles,Armstrong1967}, to imaging\cite{Campagnola2011}, material science\cite{shen1989}, and self-referencing of frequency combs\cite{Reichert1999}, to name a few.
Since its inception in 1961\cite{Franken1961}, SHG has widely been applied in bulk optics, whereas the nonlinear nature of the process requires an appropriate combination of i) a high-intensity coherent source, ii) a material endowed with a second-order nonlinearity ($\chi^{(2)}$) and iii) carefully engineered phase-matching conditions.
Such hurdles have stimulated a great research effort in the domain of integrated optics, whereas the integration of frequency doubling on-chip bears the promise of realizing novel devices in a compact, power-efficient and scalable fashion, while eliminating the need for bulky and complex experimental apparatuses.

On-chip integration has indeed proven advantageous, as it enables to enhance the interaction strength thanks to the transverse confinement in waveguides\cite{Wang2017}, and also via the use of resonant structures\cite{Wang2020}.
Moreover, the development of nano-fabrication techniques in several $\chi^{(2)}$ materials has yielded the demonstration of highly-efficient SHG in platforms such as thin film lithium niobate (TFLN)\cite{Lu2020} and III-V semiconductors, such as AlN\cite{Bruch2018}, GaN\cite{Wang2020}, AlGaAs\cite{Kuo2014} and GaP\cite{Lake2016}, especially in combination with quasi-phase-matching (QPM) techniques.
However, despite great progress, such emerging platforms struggle to find employ in practical devices, owing to the lack of compatibility with established fabrication processes, in particular those of the silicon-based complementary metal-oxide semiconductor (CMOS) technology, widely adopted by the electronics market.

In contrast,  silicon nitride (Si$_3$N$_4$) photonics\cite{Blumenthal2018, Bucio2020, Xiang2022} has emerged as a mature integrated photonics platform, thanks to its compatibility with CMOS fabrication, that favors scalability while allowing co-integration with microelectronics.
Silicon nitride photonic devices benefit notably from ultra-low propagation losses, a wide transparency window, ranging from the mid-infrared to the near-UV, large bandgap and negligible Raman effect, making them an ideal choice for high-power and specifically nonlinear applications, such as Kerr microcombs\cite{Kippenberg2018, Gaeta2019}, supercontinuum generation\cite{Grassani2019}, and parametric quantum light sources\cite{Kues2019, Arrazola2021}.
While these advantages are usually ascribed to third-order ($\chi^{(3)}$) nonlinear processes, owing to the centrosymmetric nature of the amorphous material, it was recently shown that Si$_3$N$_4$ waveguides\cite{Billat2017,Porcel2017,Hickstein2019} and resonators\cite{Lu2021,Nitiss2022} can be endowed with a photoinduced second-order nonlinearity by the coherent photogalvanic effect\added{.
This phenomenon leverages the generation of coherent currents, driven by interference between single- and multi-photon absorption, to break the centro-symmetry of the amorphous material, resulting in the inscription of a $\chi^{(2)}$ grating which, by reflecting the phase difference between the propagating fields, automatically satisfies the QPM condition.}
\replaced{This technique proved to grant }{whereas} high conversion efficiency of second harmonic \added{(SH)} light \deleted{(}$\mathrm{CE}=P_{\rm SH}/P_{\rm FH}^2$\added{(where $P_{\rm FH}$ and $P_{\rm SH}$ denote respectively the fundamental (FH) and SH power}), exceeding 2,500\%/W, \added{that} can be reached particularly in resonant structures, owing to the high values of field enhancement achievable.

Furthermore, the lack of active sources in group IV semiconductors can be overcome by heterogeneous integration with III-V sources\cite{Xiang2021integrated}.
As shown by recent findings, the combination of such sources with silicon nitride chips is not only technologically accessible, but can also be exploited to improve the coherence properties of the latter through a self-injection-locking (SIL) mechanism to a microring resonator\cite{Dahmani1987, Kondratiev2017, Raja2019, Xiang2021soliton, Lihachev2021, Jin2021, Corato-Zanarella2023, Kondratiev2023}.
In these schemes, the backscattered light from a high quality factor (Q) microring resonator is injected into the cavity of a semiconductor laser.
\added{This phenomenon is likely to naturally occur in standard microring resonators, owing to the Rayleigh backscattering due to sidewall roughness, and results in a small fraction of the input light to be back-injected in the laser cavity.}
Under appropriate \added{phase and detuning} conditions, the ring resonator acts as an effective narrowband filter, resulting in the locking of the source to its resonance frequency and a \added{significant} narrowing of the laser linewidth, \replaced{which scales proportionally }{proportional} to $Q^2$. 
Recent experimental evidence has shown a reduction of the intrinsic laser linewidth below the hertz-level\cite{Jin2021}, while the combination of this technique with Kerr microcombs has led to the realization of heterogeneously integrated turnkey soliton sources\cite{Shen2020,Xiang2021soliton}.

\begin{figure*}[ht!]
    \centering
    \includegraphics{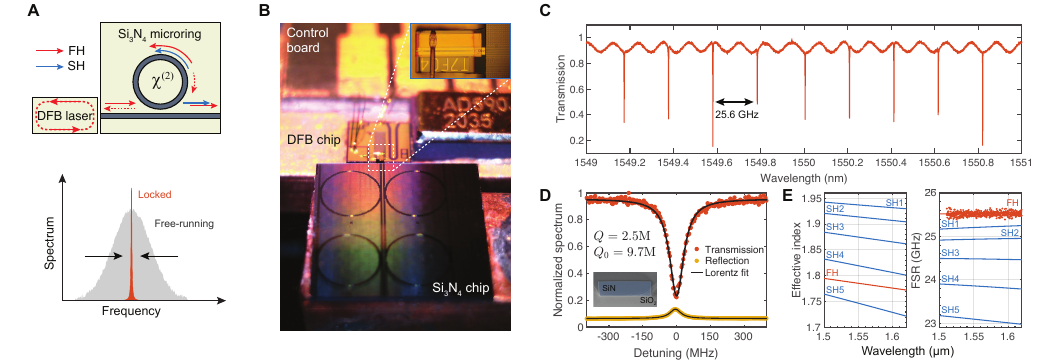}
    \caption{
        \textbf{Self-injection-locked second-harmonic source.} 
        \textbf{a.} Schematic of the SIL mechanism. 
        The DFB laser injects light at the FH wavelength (solid red arrow) to the ring resonator bus waveguide.
        A small fraction of the light circulating inside the ring is reflected by Rayleigh backscattering (dashed arrows) and injected  to the DFB cavity, yielding a dramatic narrowing of the emission linewidth compared to the free-running regime (lower panel).
        Such high-coherence laser field displays a high intracavity intensity, which is used to trigger the coherent photogalvanic effect and generate SH light (blue arrow). \added{Note that the backscattered SH light is not shown as it does not participate to the SIL process.}
        \textbf{b.} Schematic of the experimental device.
        The DFB laser (detail in inset) is mounted and wire-bonded on a temperature stabilized control board, whose position is finely tuned by micro-actuators.
        The temperature of the Si$_3$N$_4$ chip is independently controlled to tune the resonant conditions.
        Light is collected at the output using a collimation lens or a lensed fiber (not shown here).
        The Si$_3$N$_4$ chip size is $5\times5$ mm$^2$.
        The DFB chip length is \SI{400}{\micro\meter}.
        \textbf{c.} Transmission spectrum at the the FH wavelength.
        \textbf{d.} Detail of a resonance used for SIL-SHG.
        The inset shows a SEM cross-section of one of the fabricated waveguides.
        \textbf{e.} Finite element simulation of the dispersion for the TE modes involved at the FH and SH frequencies (solid lines).
        Dots are experimental data.
        The horizontal axis refers to the pump wavelength. 
    }
    \label{fig:fig1}
\end{figure*}

In this work, we show how SIL and photoinduced second-order nonlinearity can occur concurrently in a  Si$_3$N$_4$ microring resonator to create a standalone dual-wavelength source emitting highly-coherent light at both the \replaced{FH}{fundamental} and \replaced{SH}{second harmonic (SH)} frequencies (Fig. \ref{fig:fig1}a). 
The resonator's high Q factor yields a narrowing of the intrinsic laser linewidth near the hertz-level, and its high finesse results in the efficient generation of milliwatt-level light at the SH, despite the use of a non-intrinsic $\chi^{(2)}$ material.
We show how the generated SH wavelength can be tuned by simply adjusting the device operating conditions (current and temperature), and we provide a full mapping of the operating points over the C and L bands, showing abundance of suitable doubly-resonant conditions.
Remarkably, the generated light shares the same properties of the pump field, including its coherence.
This establishes our chip-scale source as a potentially powerful tool for applications that benefit not only from the ultra-narrow linewidth, such as Rb\cite{Newman2019} and Sr\cite{Origlia2016} based chip-scale atomic clocks and integrated quantum photonics\cite{Mehta2020,Moody2022}, but also from mutual coherence of the two beams, such as the self-referencing of optical frequency combs\cite{Reichert1999}.

\section*{Results}
\noindent A prototype realization of our device is shown in Fig. \ref{fig:fig1}b.
The layout consists of an electrically pumped distributed feedback laser diode (DFB), edge-coupled to the Si$_3$N$_4$ photonic chip.
The DFB is realized in an InGaAsP multi-quantum well, buried waveguide geometry, and it is packaged on a wire-bonded and thermally stabilized stage.
To prove the generality of our results, in this study we used two batches of DFB lasers, operating respectively in the C and L telecom bands and characterized by similar performance and output power, respectively up to 60~mW and 90~mW at room temperature.
By varying the driving current and device temperature, the emission wavelength can be tuned within a range of approximately \SI{5}{\nm}.
The Si$_3$N$_4$ chip is fabricated at wafer-scale through the photonic Damascene process \cite{Liu2021} and contains microring resonators with a radius of \SI{900}{\micro\meter}.
The waveguide nominal cross-section is $2\times0.55$ \si{\micro\meter\squared}, and adiabatic mode converters are used for input- and output-coupling.
Also in this case, several chips from the same wafer were tested in order to assess repeatability, with the same nominal  geometric parameters.
In this demonstration, the DFB position is finely adjusted to inject light efficiently to the bus waveguide. 
Light is then coupled to the microring resonator through a single point coupler with a gap of \SI{550}{\nano\meter}. 
We measure an insertion loss of approximately 5~dB at the DFB/Si$_3$N$_4$ chip interface \added{(see Methods)}.
The temperature of the Si$_3$N$_4$ chip is separately controlled to finely tune the resonant frequencies through thermo-optic effect.
Finally, light at the output facet is collected either in free-space or by means of a lensed fiber.
A full schematic of the experimental setup is shown in Supplementary Note 1.

Before operating in SIL regime, we characterized both the linear and nonlinear properties of the resonator in the telecom band (1510-1620 nm) using a table-top external cavity tunable laser source.
Transmission spectroscopy (Fig. \ref{fig:fig1}c) at low power reveals a set of slightly overcoupled resonances with an average free spectral range (FSR) of 25.6 GHz for the fundamental transverse electric (TE$_{00}$) mode -- in good agreement with simulations (Fig. \ref{fig:fig1}e) -- a loaded quality factor of $Q=2.5\times10^6$, and an estimated intrinsic quality factor of $Q_0=9.7\times10^6$ (Fig. \ref{fig:fig1}d).
An analogous narrowband feature is observed in the backreflected signal in correspondence with the transmission dips, which we attribute to cavity-enhanced Rayleigh backscattering inside the microresonator.
In the SH band (750-800 nm), both the ring and the bus waveguides display a multimode behavior, with up to 5 TE modes supported, which we hereby label from SH1 (TE$_{00}$) to SH5 (TE$_{40}$). The associated azimuthal resonances are undercoupled to the fundamental mode of the bus waveguide, and therefore not accessible through standard transmission spectroscopy.
However, their FSR can be accurately estimated via numerical simulations (Fig. \ref{fig:fig1}e).

\begin{figure*}[ht!]
    \centering
    \includegraphics{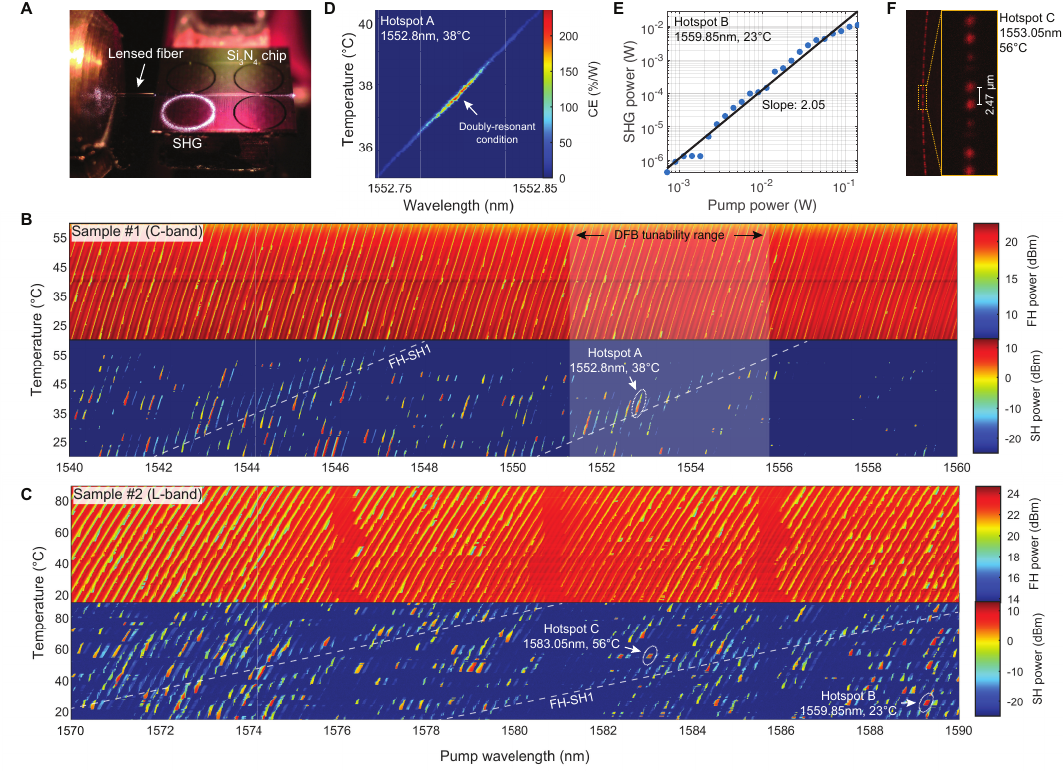}
    \caption{
    \textbf{All-optical poling and second harmonic generation.}
    \textbf{a.} Ring resonator all-optical poling using an external tunable laser.
    Light is in-coupled using a lensed fiber.
    The generated SH light is visible to the camera sensor showing the intense circulating power inside the resonator.
    \textbf{b-c.} Two dimensional maps of the output FH and SH obtained by scanning the pump power at different temperatures of the sample.
    Panel b shows the results for a Si$_3$N$_4$ sample designed to operate in the C band, while panel c corresponds to a different sample operated in the L band.
    The approximate tunability region of one of the DFBs is shaded in white for illustration.
    The dashed lines are guides to the eye highlighting the variation of the doubly-resonant condition for the corresponding families of modes.
    \textbf{d.} Map of the CE as a function of the pump wavelength and sample temperature, highlighting the best detuning condition.
    \textbf{e.} Scaling trend of the generated peak SH power as a function of the input pump level.
    \textbf{f.} Two-photon microscope image of the inscribed $\chi^{(2)}$ grating.
    }
    \label{fig:fig2}
\end{figure*}

The same tunable laser was then amplified and used to pump the resonant modes in order to investigate and map the second-order nonlinear response (Fig. \ref{fig:fig2}a).
Owing to the high field enhancement provided by resonant modes at the SH frequency, the CE is maximized whenever a nearly doubly-resonant condition is met, that is, as the pump and SH frequencies are both tuned closely to a resonance\cite{Nitiss2022}.
When this condition is satisfied and the circulating pump intensity is high enough, the coherent photogalvanic effect is triggered: a static electric field is established inside the waveguide due to the displacement of heavy charges, resulting in the breaking of the centro-symmetry condition, and in the consequent establishment of a permanently photoinduced $\chi^{(2)}$ response\cite{Dianov1995, Yakar2022}.
Moreover, the local sign and amplitude of the $\chi^{(2)}$ is such that the QPM condition is automatically fulfilled, resulting in the inscription of a $\chi^{(2)}$ grating inside the waveguide\cite{Nitiss2020formation,Nitiss2022}.
Such all-optical poling (AOP) phenomenon manifests in the sudden increase of the generated SH signal, which reaches its equilibrium state in the millisecond timescale, as soon as the appropriate pump detuning and power conditions are met.
Since the occurrence of a doubly-resonant condition is strongly dependent on the fabrication tolerance, we implemented a technique to map the AOP-SHG configurations displaying the highest CE as a function of the sample temperature and pump wavelength.
The results are shown in Figs. \ref{fig:fig2}b-c, where two different samples -- targeting respectively the C and L bands -- were measured by slowly scanning the pump laser (scan speed: 50~pm/s) at varying values of the sample temperature.
Such two-dimensional maps reveal the presence of families of doubly-resonant configurations, that can be visually identified as linear patterns in the generated SH plots (highlighted by the white dashed lines).
The slope of such patterns depends on both the FSR difference between the two modes involved and on their thermo-optic coefficient\cite{Nitiss2023}, while their horizontal spacing depends only on the FSR difference (see Methods).
From comparison between such estimated slopes and the calculated group index ad different wavelength, we are able to retrieve the resonant modes involved, in this case the pair FH-SH1.
Notably, the presence of \added{``}hotspots\added{''} \added{(namely combinations of temperature and pump wavelength characterized by high CE)} associated with particularly high generated power -- exceeding 20~mW -- was observed, which we attribute to fluctuations in the resonances Q factor and coupling conditions.
These hotspots -- some of which, falling within the DFBs tunability range, we labeled from A to C for illustrating purpose (additional data are shown in Supplementary Note 2) -- represent the optimal operating points for self-injection-locked SHG, and we therefore investigated further their properties.
To confirm the validity of our picture, we applied our mapping technique in a narrow range after optical poling with significantly lower input power (14~mW) and higher scan speed (1~nm/s), in order not to alter the properties of the inscribed grating.
The result at hotspot A, shown in Fig. \ref{fig:fig2}d, confirms the existence of an optimal combination of parameters in the temperature/wavelength space.
Furthermore, remarkably, it points to a significant increase of the CE compared to the high-power case, which we registered to be as high as 250\%/W.
This increase in the CE at low power is due to the absence of parasitic effects such as pump depletion and the generation of free-carriers associated to the nonlinear photoconductivity\cite{Yakar2022}.
A similar value of generated power and CE can be assessed for several of the hotspots identified, both in the C and L bands, that were also observed to preserve a high CE up to several tens of milliwatts of pump power.
This was confirmed by power scaling measurements, illustrated for hotspot B in Fig. \ref{fig:fig2}e, which display a nearly quadratic trend as a function of the input pump power up to about 50~mW.
Finally, to confirm the quasi-phase-matched nature of SHG, we performed two-photon imaging of the $\chi^{(2)}$ grating\cite{Nitiss2020formation} after poling the sample at hotspot C.
The result, shown in Fig.~\ref{fig:fig2}f reveals a periodic pattern, with a period of approximately \SI{2.47}{\micro\meter}.
From comparison with the simulated values of the effective index (see Methods), we infer a QPM condition between the FH and SH1 mode, in excellent agreement with our deductions drawn from the linear pattern observed in Fig. \ref{fig:fig2}c.

\begin{figure*}[ht!]
    \centering
    \includegraphics{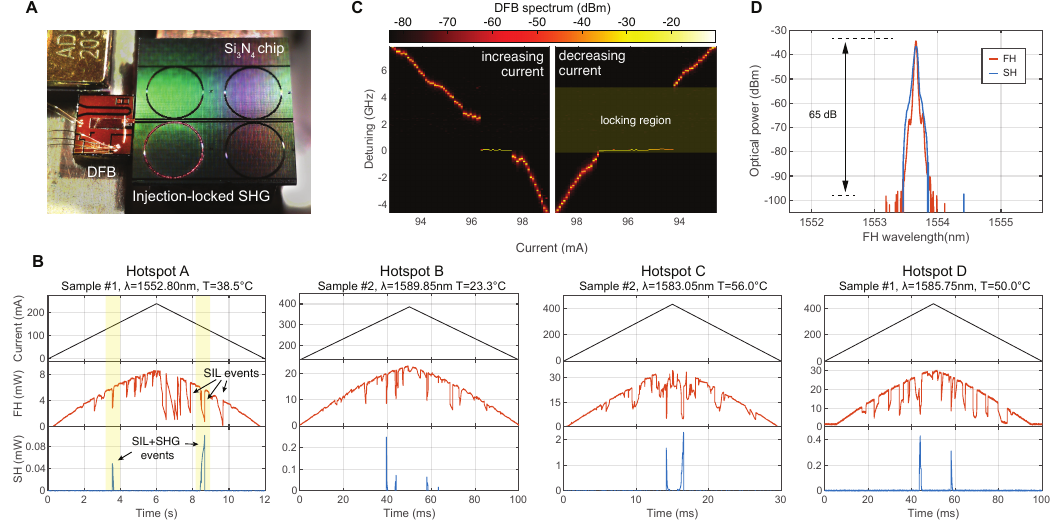}
    \caption{
    \textbf{Self-injection-locked second-harmonic generation.}
    \textbf{a.} 
    The SIL-SHG source in operation.
    The generated SH is visible as scattering from the silicon nitride ring and at the chip output.
    \textbf{b.}
    Current scans of the DFB laser targeting several operating points (hotspots) among two fo the samples studied (an AOP map including hotspot D is shown in Supplementary Note 2).
    The SIL events are identified as strongly asymmetric dips characterized by hysteresis, whose width and depth depend on the amplitude and phase of the backscattered signal by the microresonator, as well as the dip visibility.
    SIL-SHG events are marked by spikes in the generated SH power, that reaches a peak value of 2.3~mW.
    All powers are rated at the output of the bus waveguide.
    \textbf{c.}
    Optical heterodyne spectrum of the output FH as a function of the driving current.
    In the proximity of the resonance, the emission frequency deviates from the otherwise linear shifting trends, and the linewidth is narrowed significantly.
    Once the current if further increased (or decreased), the trend recovers its linear character.
    The height of the gap in the frequency range spanned by the laser emission represents the locking bandwidth.
    \textbf{d.}
    Device emission at the FH and SH wavelength recorded by an optical spectrum analyzer (resolution: 20 pm), showing the absence of side modes within the instrument's dynamic range (65~dB).
    }
    \label{fig:fig3}
\end{figure*}

From the measurements shown in Fig. \ref{fig:fig2}b-c, we identified several operating points compatible with the tunability bandwidth of our DFBs, that could be probed for self-injection-locked SHG.
The tunable laser was thus replaced by the DFB diode, realizing the SIL-SHG source as described above (Fig. \ref{fig:fig3}a).
By fixing the DFB temperature and sweeping the driving current, we were able to observe several SIL events, marked by strongly asymmetric dips in the transmitted spectrum (Fig.~\ref{fig:fig3}b), owing to the locking of the lasing frequency to the bottom of the resonance dip.
The width of such dips reflects the locking bandwidth, which depends on both the magnitude of the backreflected signal and on its phase, two quantities that are controlled by finely adjusting the position of the DFB with respect to the chip facet using a piezoelectric positioner.
By retrieving the DFB spectrum as a function of the driving current through optical heterodyne measurements (Fig. \ref{fig:fig3}c), we estimated a locking bandwidth in the GHz range, visualized by stark deviation in the otherwise linear frequency shifting trend, and characterized by a pronounced hysteresis\cite{Kondratiev2017} between the increasing current (decreasing frequency) and the decreasing current (increasing frequency) scans.
When the doubly-resonant condition is fulfilled in correspondence of a SIL event, the AOP effect is triggered, resulting in the emission of light at the SH frequency.
The phenomenon was investigated for several doubly-resonant configurations (hotspots) shown in Figs. \ref{fig:fig2}b-c and the result was found consistently repeatable both for pumping in the C and L bands, with a maximum emission power as high as 2.3~mW in the bus waveguide, and a peak CE as high as 280\%/W\added{, corresponding to a nonlinearity susceptibility $\chi^{(2)}=\SI{0.54}{\pico\meter\per\volt}$ (see Supplementary Note 3), consistent with previously reported values\cite{Porcel2017, Billat2017, Hickstein2019, Grassani2019, Nitiss2020formation, Lu2020, Nitiss2022}}.
Remarkably, the former value is comparable, if not greater, to the best results obtained in for self-injection-locking in TFLN technology\cite{Ling2023}.
This result highlights the high potential of silicon nitride for the engineering of second-order nonlinear processes, despite relying on a photoinduced, rather than intrinsic, nonlinearity.
By fixing the current in correspondence of a SIL-SHG event, a constant CW emission is observed (Fig.~\ref{fig:fig3}a).
When visualized on an optical spectrum analyzer (Fig.~\ref{fig:fig3}d), such emission shows a monochromatic spectrum, characterized by strong side-mode suppression ratio (SMSR) exceeding 60~dB, limited by the sensitivity of our instrument.

\begin{figure*}[ht!]
    \centering
    \includegraphics{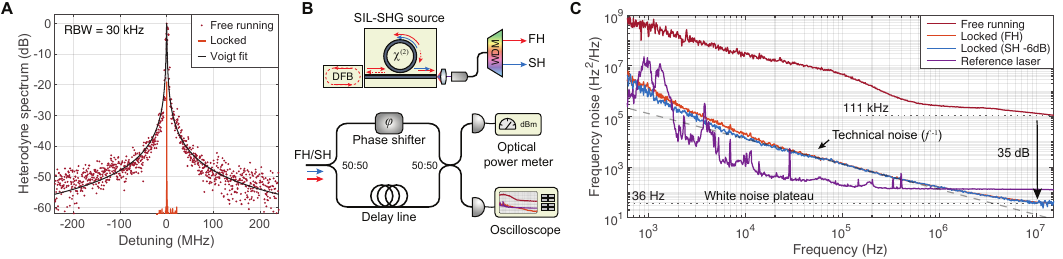}
    \caption{
    \textbf{Emission linewidth.}
    \textbf{a.}
    Heterodyne spectrum of the emission linewidth at FH in the free running (red dots) and SIL (orange line) regimes.
    The linewidth of the FH emission in the SIL regime is estimated to be well-below the resolution bandwidth (RBW) of the instrument.
    \textbf{b.}
    Schematic of the frequency discriminator setup used for the frequency noise measurements.
    Light from the SIL-SHG source is collected at the output of the chip, and the de-multiplexed FH is routed to an unbalanced Mach-Zehnder interferometer.
    On one arm, the phase is regulated using a fiber phase shifter.
    The output signal is detected at one of the outputs with a fast photodiode and recorded by an oscilloscope.
    \textbf{c.}
    Frequency noise spectra measured by the frequency discriminator.
    The technical noise pattern, scaling as $f^{-1}$ and associated with a Gaussian contribution to the broadening of the emission line, is marked by a dashed grey line.
    The white noise plateaus, associated with the Lorentzian contribution to the broadening of the emission line, are highlighted by dotted grey lines.
    From the value of such plateaus we estimated a narrowing of the intrinsic linewidth of 35~dB in the present case.
    }
    \label{fig:fig4}
\end{figure*}

Finally, we investigated the coherence properties of our dual-wavelength source.
A first analysis of the emission linewidth was performed at the fundamental wavelength by optical heterodyne with a reference tunable laser (Fig. \ref{fig:fig4}a).
A significant narrowing can be immediately appreciated when passing from the free-running to the injection-locked regime, with a decrease in linewidth from $\delta\nu \approx \SI{1}{\mega\hertz}$ in the former case (3~dB bandwidth from Voigt fit), to $\delta\nu < \SI{50}{\kilo\hertz}$ when locked, limited by the noise of the reference laser used as a local oscillator.
To get more insight on the emitted light properties, we implemented a frequency discriminator apparatus to assess the frequency noise of the emitted light\cite{Llopis2011,Tran2019}, as shown in Fig. \ref{fig:fig4}b, operating simultaneously at the FH and SH wavelengths (see Methods).
The retrieved power spectral density (PSD) for the FH and SH fields is shown in Fig. \ref{fig:fig4}c, where a dramatic reduction in both the technical ($f^{-1}$) and white noise at the FH, exceeding 35~dB, is observed, with a noise floor lower plateau as small as $S_\nu^0 = \SI{36}{\hertz\squared\per\hertz}$, corresponding to an intrinsic (Schawlow-Townes) linewidth $\delta\nu_{\rm ST}=\pi S_\nu^0=\SI{113}{\hertz}$.
The same trend, up to an offset of 6~dB, inherent to frequency doubling, is observed in the SH signal over more than 4 decades, proving the correlated nature of the FH and SH fields.
 We point out that assessing the frequency noise for FH and SH simultaneously limits the sensitivity of the measurement, which requires sufficiently high power to overcome the thermal noise of the photodetectors.
For several SIL-SHG configurations where the collection efficiency was optimized for measuring solely the FH, a lower minimum noise floor, as small as \SI{13}{\hertz\squared\per\hertz} (instrinsic linewidth: $\delta\nu_{\rm ST}=\SI{41}{\hertz}$, noise reduction: >39~dB) could be assessed for the FH (see Supplementary Note 4).
As a comparison, the SIL source outperforms the commercial tunable laser used for characterization in terms of intrinsic linewidth.

\section*{Discussion}
\noindent
The device developed here represents a novel approach to the engineering of on-chip SH sources, whereas the resonant element not only enhances the CE, but first and foremost improves the coherence properties of the FH and SH fields through the SIL mechanism.
While this effect has been widely investigated recently for perspective application to other nonlinear processes\cite{Raja2019, Shen2020, Xiang2021soliton, Lihachev2021}, the results presented here represent one of the first demonstrations of self-injection-locked SHG on a chip\cite{Clementi2023cleo}. 

Only a single similar result has been reported so far, to the best of our knowledge, by Ling and co-workers\cite{Ling2023}.
In that work, the authors exploit TFLN technology to realize a SIL-SHG source similar to the one presented here, by leveraging the high intrinsic $\chi^{(2)}$ and electric field poling of lithium niobate.
Despite promising results, their device still suffers from some limitations inherent to the TFLN platform, most notably i) the need for electrodes used to inscribe a QPM grating, that limits the operation to a fixed design wavelength, and ii) a relatively low quality factor ($Q\approx4\times10^5$), that sets the best narrowing performance reported to an estimated intrinsic linewidth of \SI{4.7}{\kilo\hertz} at the SH.
In contrast, our device displays wide tunability across the whole telecom spectrum, only requiring control over the pump laser wavelength and on the sample temperature.
The AOP mechanism allows indeed to erase and re-write the QPM grating by solely changing these two parameters, as long as a doubly-resonant condition is satisfied\cite{Nitiss2022}. 
As a result, the same microresonator can be dynamically reconfigured in order to match a different pump wavelength and/or family of modes, with  ample choice provided by the abundance of doubly-resonant configurations, thus eliminating the need for poling electrodes and enhancing the flexibility of the final device.
Our solution also excels in terms of coherence, as it displays an intrinsic linewidth almost reaching the hertz-level, owing to the high Q of the resonators used.
It is worth stressing  that such short-term linewidth is mapped on the generated SH field, with a predicted intrinsic SH linewidth as small as 163 Hz (note that a 4-fold increase in the frequency noise is expected as a result of frequency doubling), thus implying the mutual coherence between the output fields at the FH and SH wavelengths.
Our device also performs well in terms of generated power, being capable to reach and exceed a milliwatt-level SH output (up to 2.3~mW) with a pump power of about 33~mW, corresponding to a net (i.e. non-normalized) conversion efficiency $\eta=P_{\rm SH}/P_{\rm FH}\approx7\%$ at continuous-wave regime, and consistently displaying a normalized CE exceeding 100~\%/W across all the hotspots tested, with a peak value recorded as high as 280~\%/W.
This result is particularly remarkable given the relatively low value of the photoinduced nonlinearity in silicon nitride -- \replaced{here $\chi^{(2)}\approx0.54$ pm/V}{ up to $\chi^{(2)}\approx0.3$ pm/V}, compared to $\chi^{(2)}\approx54$ pm/V in the case of TFLN\cite{Wang2017} -- and highlights the maturity of the silicon nitride photonics platform, as well as its suitability for applications in nonlinear optics.
Our device performance proves superior also compared with existing single-wavelength SIL sources in the visible and near-infrared ranges\cite{Siddharth2022, Corato-Zanarella2023}, with an order-of-magnitude narrower intrinsic linewidth and significantly higher SMSR.
This advantage can be attributed to the more difficult realization of the laser and microresonator components, which requires an obvious increase in the fabrication accuracy to efficiently operate at a shorter wavelength.

In the perspective of commercial applications, our proof-of-concept realization could be further improved, both in terms of device engineering and figures of merit.
In particular, the output power can be significantly increased by engineering an optimal coupling between the DFB facet and the bus waveguide, for example through the use of optimized adiabatic mode converters.
Ultimately, one could also foresee a full heterogeneous integration, which has been shown to be within reach of state-of-art fabrication technology\cite{Xiang2021integrated}, thus enabling a wafer-scale integration of this type of sources.
\replaced{Furthermore, }{Finally, further} improvements in the microresonator Q factor may enable increased conversion efficiencies and even lower laser linewidths\added{ (as both the CE and the narrowing factor scale as $Q^2$)}, potentially unlocking access to hertz-level dual-wavelength coherence on a photonic chip.
\added{This is particularly relevant in the framework of commercial SiN fabrication, where the propagation loss are still slightly higher (typically around 0.1--0.2 dB/cm) than in state-of-the-art research (we estimate approximately 4 dB/m in this work).}
This result has indeed already proven possible in single-wavelength SIL sources\cite{Jin2021}, whereas the use of very long ring resonators in a folded spiral geometry also showed promising advantage in reducing the  thermo-refractive contribution to technical noise, the latter being effectively averaged over the whole device length.
However, this approach may not be suitable for the purpose of frequency doubling, as the use of low-confinement waveguides increases the transverse mode area, ultimately weakening the nonlinear interaction.
In this perspective, the use of high-confinement waveguides based on thick silicon nitride layers\cite{Liu2021} is more advantageous, as it maximizes the nonlinear interaction.
Moreover, the use of long resonators reduces the field enhancement, ultimately setting a trade-off between coherence and conversion efficiency.
Finally, the combination of SIL and AOP can be potentially extended to further processes, such as the cascaded sum-frequency generation of the optical third-harmonic\cite{Yakar2022,Hu2022}\added{, upon appropriate optimizations of the source, such as an increase of the circulating power, engineering of a triply-resonant condition and improved light extraction at the third-harmonic wavelength}.
Not least, one could foresee, through fine-tuned dispersion engineering, to employ the same microring resonator for the generation of a self-starting soliton microcomb\cite{Shen2020}, whose frequency doubling could potentially allow access to $f-2f$ interferometry on-chip\cite{Brasch2017}.
Despite bearing high technical difficulties, this target would come with great benefit, allowing to realize a self-referenced microcomb on a single integrated photonic chip.
Such achievement would unravel the potential of optical atomic clocks in a fully integrated chip-scale photonic device, and potentially be exploited to bring hertz-level coherence over the whole near-infrared spectrum and beyond.

In conclusion, we have demonstrated a chip-scale dual-wavelength source based on the self-injection-locking of a DFB laser to a high-Q Si$_3$N$_4$ microresonator.
The device displays a near-hertz intrinsic linewidth of 41 Hz, milliwatt-level SH output power and high side-modes suppression exceeding 60~dB, over a locking bandwidth of several gigahertz.
By exploiting an all-optical poling technique, our system can be reconfigured to operate across the whole C and L telecom bands by solely tuning the sample temperature and pump wavelength.
Our findings confirm the suitability of silicon nitride photonics for the integration of highly-efficient second-order nonlinear processes, and open a pathway towards the realization of novel chip-scale devices such as miniaturized atomic clocks and fully integrated self-referenced microcombs.

\noindent \emph{Note.}
During the preparation of this manuscript, a similar example of SIL-SHG in silicon nitride has been reported online\cite{Li2023} in the form of a preprint.

\vspace{0.5cm}
\noindent\textbf{Methods}
\medskip
\begin{footnotesize}

\noindent\textbf{Laser diodes}.
The devices are distributed feedback (DFB) lasers with high-reflection coating at the rear facet and anti-reflection coating at the front facet.
The technology is a semi-insulating buried heterostructure, where the waveguide is buried into semi-insulating InP, leading to a circular optical mode, and efficient thermal dissipation.
The C-band lasers are standard \SI{400}{\micro\meter}-long DFB lasers.
The L-band lasers are specifically designed for high power with a long laser cavity of \SI{800}{\micro\meter}. 
Besides, they provide improved coupling efficiency with the integration of a spot-size converter at the output, leading to a broad circular mode.
Wavelength tuning with current is due to heating of the laser section by Joule effect, which increases effective index. 
Tuning with temperature is \SI{0.9}{\nano\meter\per\celsius}. 
Heating with injected electrical power is about \SI{50}{\celsius\per\watt} for the chips on carrier.
\added{
A typical lasing curve at \SI{20}{\celsius} is provided below, together with a correspondent output from the Si$_3$N$_4$ chip, showing a typical insertion loss of approximately \SI{5}{\decibel} at the chip facet.
This insertion loss was found to be consistent among all the devices tested.
\begin{figure}[ht!]
    \centering
    \includegraphics[width=0.3\textwidth]{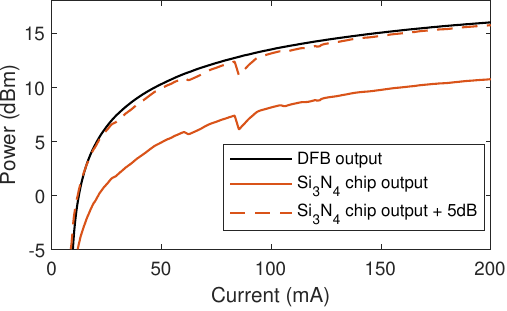}
\end{figure}
}

\vspace{0.1cm}

\noindent\textbf{Microring resonators}.
The Si$_3$N$_4$ microresonators used in the experiment were fabricated by the photonic Damascene process that yields ultralow propagation loss\cite{Liu2021}. 
They have a ring structure (radius $R=\SI{900}{\micro\meter}$) coupled with a bus waveguide buried in SiO$_2$ cladding.
The waveguide nominal cross-section (width $\times$ height) is $2000\times550~{\rm nm^2}$, which supports multiple spatial modes at the SH wavelength.
A scanning electron microscope image of the cleaved sample reveals a fabricated cross-section of $2150\times572~{\rm nm^2}$, which was used for simulation of the resonant modes.
The microresonator is characterized to exhibit normal dispersion at the pump wavelength.
\added{All the simulations shown in this work have been performed using the Ansys Lumerical software suite.
Discrepancies between the experimental and simulated data are attributed to the relative accuracy of our tunable laser source.}

\vspace{0.1cm}

\noindent\textbf{AOP mapping}.
\noindent The formula used to calculate the slope of the doubly-resonant trends in Fig. \ref{fig:fig2}b-c is\cite{Nitiss2023}:
\begin{equation}
    \frac{dT}{d\lambda_{\rm p}} \approx -\frac{\Delta\nu_{\rm FH}-\Delta\nu_{\rm SH}}{\Delta\nu_{\rm FH} \left(\frac{d\lambda_{\rm FH}}{dT}-2\frac{d\lambda_{\rm SH}}{dT}\right)}
\end{equation}
\noindent
where $T$ is the sample temperature, $\lambda_{\rm p}$ is the pump wavelength, $\Delta\nu_{\rm FH(SH)}$ is the FSR at the FH (SH) wavelength expressed in hertz and $d\lambda_{\rm FH(SH)}/dT$ is the thermo-optic coefficient at
the FH (SH) wavelength.
The approximation is valid as long as the ratio between the two FSRs is close to 1.
The horizontal spacing between similar trends is calculated as:
\begin{equation}
    \Delta\lambda_{\rm spacing} \approx \frac{\lambda_{\rm p}^2\Delta\nu_{\rm FH}^2}{2c\lvert\Delta\nu_{\rm FH}-\Delta\nu_{\rm SH}\rvert}
\end{equation}
\noindent
where $c$ is the speed of light.
It is calculated to be around \SI{8.3}{\nano\meter} around $\lambda_{\rm p}=\SI{1550}{\nano\meter}$ for the FH-SH1 mode pair.

\vspace{0.1cm}
\noindent\textbf{TPM imaging}. 
For characterization of the inscribed $\chi^{(2)}$ gratings, a high power femtosecond Ti:Sapphire laser is focused at the grating plane of the microresonator in an upright configuration. 
The focal spot is then raster-scanned across the plane while, in the meantime, its generated SH signal is monitored so that the (squared) $\chi^{(2)}$ response is probed.
From the periodicity retrieved, the original phase mismatch between the modes involved is inferred as:
\begin{equation}
\frac{2\pi}{\Lambda}=
\frac{2\omega}{c}
\lvert n_{\rm eff}^{\rm SH} - n_{\rm eff}^{\rm FH}\rvert
\end{equation}
where $\Lambda$ is the poling period, $\omega$ is the pump angular frequency and $n_{\rm eff}^{\rm FH(SH)}$ is the effective index of the FH (SH) mode.
From simulations, we estimate $\Lambda/2=\SI{2.45}{\micro\meter}$, in good agreement with the result shown in Fig.~\ref{fig:fig2}f.

\vspace{0.1cm}
\noindent\textbf{Optical heterodyne}.
To obtain the data shown in Figs.~\ref{fig:fig3}c and \ref{fig:fig4}a, an external cavity laser, serving as a local oscillator (LO) is tuned at frequency close to the emission under test.
The two fields are mixed at a 50:50 fiber beam-splitter and routed to a fast photodiode, in order to retrieve the optical beat-note.
The resulting electrical signal is visualized on an electrical spectrum analyzer, retrieving the narrowband spectrum of the emitted light.
The resolution of the technique is approximately 50 kHz, and it is limited by the resolution bandwidth of the spectrum analyzer and by the finite linewidth of the LO laser.

\vspace{0.1cm}
\noindent\textbf{Frequency discriminator}.
For the measurement of the frequency noise PSDs shown in Fig.~\ref{fig:fig4}c, we implemented two identical frequency discriminators operating respectively at the FH and SH wavelengths.
Such systems, schematized in Fig.~\ref{fig:fig4}b (see Supplementary Note 1 for more detailed schematics), consist in an unbalanced Mach-Zehnder interferometer, which is used to map phase fluctuations into amplitude fluctuations, that could be detected by a fast photodiode and recorded by an oscilloscope or an electrical spectrum analyzer.
The relative phase between the arms is scrambled using a fiber phase shifter.
During the SIL-SHG process, both the FH and SH noise are recorded simultaneously, while the average output power is recorded to calibrate the measurement.


\end{footnotesize}

\vspace{0.5cm}
\noindent\textbf{Acknowledgements}
\medskip
\begin{footnotesize}

\noindent 
This work was funded by ERC grant PISSARRO (ERC-2017-CoG 771647).

\end{footnotesize}

\vspace{0.5cm}
\noindent\textbf{Competing interests}
\medskip
\begin{footnotesize}

\noindent
The authors declare no competing interests.

\end{footnotesize}

\vspace{0.5cm}
\noindent\textbf{Data availability}
\medskip
\begin{footnotesize}

\noindent
The data and code that support the plots within this paper and other findings of this study are available from the corresponding authors upon reasonable request.

\end{footnotesize}

\bibliography{references}

\end{document}